\documentclass[preprint2]{emulateapj}

\shorttitle{Spectral LDFC}
\shortauthors{Guyon et al.}
\usepackage{graphicx}

\usepackage[mathscr]{eucal}
\usepackage{amsfonts}
\usepackage{mathtools}

\begin{document}

\title{Spectral Linear Dark Field Control: Stabilizing Deep Contrast for Exoplanet Imaging Using out-of-band Speckle Field}
\author{Olivier Guyon}

\affil{Astrobiology Center, National Institutes of Natural Sciences, 2-21-1 Osawa, Mitaka, Tokyo, JAPAN }
\affil{Steward Observatory, University of Arizona, Tucson, 933 N Cherry Ave, Tucson, AZ 85721, USA}
\affil{College of Optical Sciences, University of Arizona, 1630 E University Blvd, Tucson, AZ 85719, USA}
\affil{National Astronomical Observatory of Japan, Subaru Telescope, National Institutes of Natural Sciences, Hilo, HI 96720, USA}
\email{oliv.guyon@gmail.com}

\author{Kelsey Miller}
\affil{College of Optical Sciences, University of Arizona, 1630 E University Blvd, Tucson, AZ 85719, USA}
\affil{Steward Observatory, University of Arizona, Tucson, 933 N Cherry Ave, Tucson, AZ 85721, USA}

\author{Jared R. Males}
\affil{Steward Observatory, University of Arizona, Tucson, AZ 85721, USA}

\author{Ruslan Belikov}
\affil{NASA Ames Research Center, Moffett Blvd, Mountain View, CA 94035, USA}

\author{Brian D. Kern}
\affil{Jet Propulsion Laboratory, 4800 Oak Grove Dr, Pasadena, CA 91109, USA}

\begin{abstract}
Wavefront stabilization is a fundamental challenge to high contrast imaging of exoplanets. For both space and ground observations, wavefront control performance is ultimately limited by the finite amount of starlight available for sensing, so wavefront measurements must be as efficient as possible. To meet this challenge, we propose to sense residual errors using bright focal-plane speckles at wavelengths outside the high contrast spectral bandwidth. We show that a linear relationship exists between the intensity of the bright out-of-band speckles and residual wavefront aberrations. An efficient linear control loop can exploit this relationship. The proposed scheme, referred to as Spectral Linear Dark Field Control (spectral LDFC), is more sensitive than conventional approaches for ultra-high contrast imaging. Spectral LDFC is closely related to, and can be combined with, the recently proposed spatial LDFC which uses light at the observation wavelength but located outside of the high contrast area in the focal plane image. Both LDFC techniques do not require starlight to be mixed with the high contrast speckle field, so full-sensitivity uninterrupted high contrast observations can be conducted simultaneously with wavefront correction iterations. We also show that LDFC is robust against deformable mirror calibration errors and drifts, as it relies on detector response stability instead of deformable mirror stability. LDFC is particularly advantageous when science acquisition is performed at a non-optimal wavefront sensing wavelength, such as nearIR observations of planets around solar-type stars, for which visible-light speckle sensing is ideal.

We describe the approach at a fundamental level and provide an algorithm for its implementation. We demonstrate, through numerical simulation, that spectral LDFC is well-suited for picometer-level cophasing of a large segmented space telescope. 
\end{abstract}


\section{INTRODUCTION}
\label{sec:intro}

\begin{figure*}[htb]
\includegraphics[scale=0.3]{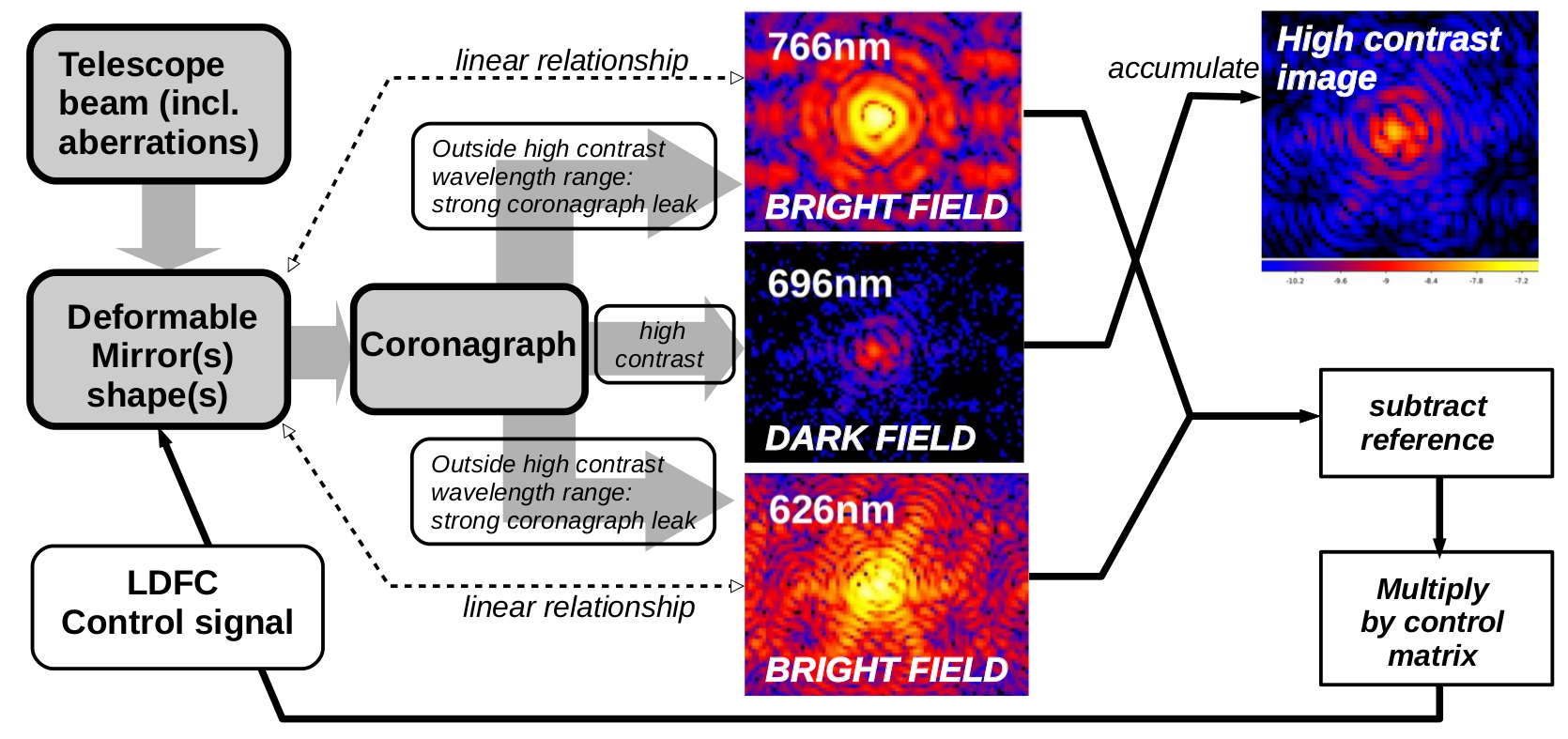} 
\caption[LDFCprincip] { \label{fig:LDFCprincip} Spectral Linear Dark Field Control (LDFC) principle. In this example, an image is acquired on each side of the high contrast wavelength (3 images total). The bright field images (766nm and 626nm) drive a linear wavefront contol loop. Individual images shown in this figure are monochromatic.}
\end{figure*}

Direct imaging and spectroscopic characterization of habitable exoplanets requires high contrast imaging capability, which is achieved by combining a coronagraph and a high performance wavefront control system \citep{1995PASP..107..386M, 2009arXiv0911.3200L}. Wavefront stabilization is the most significant fundamental challenge to this endeavour \citep{2005SPIE.5905..110S}. An Earth-like planet orbiting a Sun-like star is approximately 1.5e-10 times as bright as the star, and at $\lambda = $ 500nm, a pupil plane sine wave aberration of 2 picometer amplitude is sufficient to create an equally bright speckle. The wavefront stabilization challenge is a tradeoff between sensing sensitivity and optical stability \citep{2005ApJ...629..592G}: wavefront variations occuring within the time required for sensing cannot be temporally resolved and are left uncorrected.

To meet this challenge, imaging systems operating at high contrast currently use focal-plane wavefront sensing and control \citep{2006ApJ...638..488B, 2009SPIE.7440E..0DG, 2016A&A...593A..33M, 2017MNRAS.464.2937B}: unwanted speckles in the focal plane are probed by a known set of deformable mirror (DM) actuations so that their amplitude and phase can be recovered and then canceled by adding with the same DM(s) speckles of opposite complex amplitude. This process, referred to as Electric Field Conjugation (EFC), creates in the image a dark field (DF) suitable for high-contrast imaging. Measurement cycles must be repeated sufficiently frequently to track wavefront changes, and can be time-consuming due to the small amount of residual starlight available in the DF. The corresponding probed images may also not be scientifically useful due to the added starlight component, and precious observing time must be shared between wavefront sensing and scientific integration.

We describe in this paper an efficient alternative to this process, locking the wavefront using bright speckles located outside the DF in both spatial and spectral dimensions. As described in \cite{2017arXiv170304259M}, intensity changes in these bright field (BF) regions are linearly related to the same wavefront changes that spoil the deep halo suppression in the DF. Because the BF images are significantly brighter than the DF images, they can be acquired at higher cadence, and no additional starlight needs to be directed to the DF during science exposures. By calibrating or computing the linear changes in the BF against wavefront changes, a linear dark field control (LDFC) servo can maintain high-contrast in the DF during science exposures. In this paper, we extend the spatial LDFC technique presented in \cite{2017arXiv170304259M} into the spectral domain, and demonstrate close-loop wavefront control for a simulated high-performance coronagraph operating on a segmented aperture. The spectral LDFC principle and algorithm steps are detailed in \S \ref{sec:princip}. We describe and quantify its sensitivity benefits in \S \ref{sec:gain}. Other benfits are discussed in \S \ref{sec:imsensingbenefits}. We demonstrate in \S \ref{sec:numsimu} picometer-level cophasing of a large segmented aperture for high contrast imaging. Implementation challenges, limitations and future capabilities are discussed in \S \ref{sec:disc}.

\section{Principle}
\label{sec:princip}

\subsection{Spectral Linear Dark Field Control (LDFC) overview}

In Linear Dark Field Control (LDFC), starlight falling outside the high contrast region is used to drive a linear control loop aimed at “freezing” the wavefront state. This bright light (referred to as the bright field, BF) is abundant relative to the much smaller amount of light in the high contrast region (referred to as the dark field, DF). In the small perturbation regime, BF intensity responds linearly to wavefront changes, allowing a linear control loop to operate from BF intensity (sensor) to wavefront control commands. In {\bf spatial LDFC}, the BF used for sensing consists of light that is spatially outside the high contrast imaging area, as described in \cite{2017arXiv170304259M}. We study in this paper {\bf spectral LDFC}, where sensing relies on BF light that is at wavelengths on either side of the high contrast spectral band, as illustrated in Fig. \ref{fig:LDFCprincip}. Readers can refer to \cite{2017arXiv170304259M} for a derivation of the linear relationship between BF intensity and wavefront complex amplitude, noting that it applies equally to spatial and spectral LDFC. 

Coronagraph masks and wavefront control systems are typically optimized to deliver high contrast in a $\approx$ 10-20\% wide spectral band \citep{2016JATIS...2a1013T}. The amount of residual starlight in the post-coronagraph focal plane image increases rapidly outside this band, providing the light required for spectral LDFC. Broader spectral band images of exoplanets can be assembled by combining multiple observations, obtained sequentially by changing filters, focal plane masks and DM(s) states. At any given time during this process, light outside the high contrast spectral band may be used for LDFC. 

The LDFC loop seeks to drive the BF to a pre-recorded (or pre-computed) reference BF image, so it can stabilize an already established system state delivering high contrast performance in the DF. A separate wavefront contol technique is required to initially reach the deep contrast state that will serve as the reference point for LDFC.

\subsection{Algorithm Description}
\label{ssec:algo}

Spatial or spectral LDFC is implemented as a conventional linear control loop in a high contrast imaging system, with the BF intensity image serving as the linear input sensor, and the deformable mirror(s) commands as the output control. Calibration or knowledge of the linear derivative of the BF intensity image against the deformable mirror commands is therefore required prior to closed loop operation.

The main steps for the LDFC algorithm are:
\begin{enumerate}
\item{SETUP: The system is driven to a high contrast state in the DF.}
\item{CALIBRATION: A reference BF intensity image $I_{ref}$ is acquired.}
\item{CALIBRATION: For each control actuator (of a DM), the derivative of the BF image against the actuator position is measured or computed.}
\item{CALIBRATION: Derivatives are assembled in a system response matrix (RM) linearly linking DM commands to the BF image intensity}
\item{CALIBRATION: A control matrix (CM) is computed by regularized inversion of the RM.}
\item{CLOSED LOOP OPERATION: A new BF image I is acquired.}
\item{CLOSED LOOP OPERATION: The command is computed as $-{CM} \times (I-I_{ref})$ and applied to the DM(s).}
\end{enumerate}

Steps 1-5 (setup and calibration) are performed prior to multiple iterations of steps 6-7 (closed loop control). The usual refinements of linear control algorithms, such as modal gain and predictive control, can be added to the control scheme. The proposed linear wavefront control scheme is similar to control loops in ground-based adaptive optics using pupil-plane sensors; the only difference being that focal plane pixels are used instead. The approach is closely related to, and can be considered as an extension of, the coronagraphic low-order wavefront sensor (CLOWFS) technique, which also uses light rejected by the coronagraph for high speed sensing and control \citep{2009ApJ...693...75G, 2014PASP..126..586S, 2015PASP..127..857S, 2016JATIS...2a1021S}. While the CLOWFS has little leverage on mid-spatial frequencies (to which CLOWFS is not sensitive), LDFC extends efficient linear control to mid-spatial frequencies for contrast stabilization.

\section{Sensitivity gain}
\label{sec:gain}

A core motivation behind spectral LDFC loop is to operate the high contrast wavefront control loop at higher speed and sensitivity than possible with conventional wavefront control schemes relying entirely on measurements in the high contrast spectral band. We assume in this section photon-noise limited detectors, and quantify wavefront sensing efficiency gains. The {\bf sensing efficiency} is defined here as the inverse of the time required to reach a given level of noise, so a 2x increase in efficiency means that the sensing time at fixed noise level has been halved, or, equivalently, that the measurement noise (standard deviation) at fixed exposure time is reduced by $\sqrt{2}$.

\subsection{Coherent Mixing and Duty Cycle}

Both DH speckle modulation and LDFC fundamentally rely on coherent mixing of speckles created by wavefront aberrations with starlight speckles (either existing BF light or speckles created by DM probes) for measurement. In spectral LDFC, small wavefront errors create speckles that interfere coherently in the focal plane with much brighter starlight originating from coronagraph leaks. We discuss in this section the sensitivity of this highly unbalanced coherent mixing, which occurs at wavelengths outide the high contrast spectral band.

Without loss of generality, we consider here the simplified case where the unknown focal plane complex amplitude $A$ due to wavefront aberrations is a real number (1-D simplification to the full 2-D complex amplitude plane). We assume that two intensity measurements are performed, corresponding to probes (or BF speckles for LDFC) adding respectively $\delta A$ and $- \delta A$. In spectal LDFC, these two probes may correspond to two separate wavelength channels: one for which the interference with underlying starlight is constructive, the other for which it is destructive.
\begin{equation}
I_1 = (A + \delta A)^2 \pm \sigma_1
\end{equation}
\begin{equation}
I_2 = (A - \delta A)^2 \pm \sigma_2
\end{equation}

In the photon-noise regime, measurement noises are:
\begin{equation}
\sigma_1 = |A + \delta A|
\end{equation}
\begin{equation}
\sigma_2 = |A - \delta A|
\end{equation}

The complex amplitude $A$ is estimated by differencing $I_1$ and $I_2$:
\begin{equation}
A = \frac{I_1 - I_2}{4 \: \delta A} \pm \sigma_A
\end{equation}

The associated measurement noise is:
\begin{equation}
\sigma_A = \frac{\sqrt{\sigma_1^2 + \sigma_2^2}}{4 \: \delta A} = \frac{\sqrt{A^2 + \delta A^2}}{4 \: \delta A} = \frac{\sqrt{1+x^2}}{4 x}
\end{equation}
where $x=\delta A/A$ is the ratio of probe amplitude to speckle amplitude. In LDFC, $x$ is usually very large, as the existing bright starlight acts as the probes. The smallest noise is achieved when probe amplitude is larger than speckle amplitude ($x>>1$), corresponding to the asymptotic limit $\sigma_A = 0.25$. With equal speckle and probe amplitudes ($x=1$), variance is twice as large ($\sigma_A = 0.25 \sqrt{2}$), requiring twice as much exposure time to reach the same estimation quality (in the photon-noise regime considered here, measurement variance is inverse proportional to the square root of number of photons). The conclusion also applies to the full complex amplitude (2-D) case where $A$ is a complex number, provided that the number of probes is increased to at least 3 (see \cite{2005ApJ...629..592G} for details).

By using bright speckles to measure wavefront changes, LDFC operates near the asymptotic $x >> 1$ high-flux limit, providing a 2x gain in sensitivity over a speckle modulation loop where the modulation amplitude is comparable to the speckle amplitude. DF speckle modulation can also operate at higher efficiency by adopting large amplitude probes, but the corresponding images are then no longer contributing to the science integration due to excess photon noise. By avoiding the tradeoff between sensing sensitivity and duty cycle, LDFC brings a $\approx$ 2x gain in sensing efficiency (this gain applies equally to spectral, spatial and spectral+spatial LDFC).

\subsection{Spectral Bandwidth}

Additionally, spectral LDFC allows light to be gathered over a wider spectral band than the coronagraph's designed high contrast spectral bandwidth. Coronagraphs are often optimized to deliver high contrast over a approximately 10 to 20 \% spectral bandwidth, yet spectral LDFC can gather light over approximately 40 to 80 \% spectral bandwidth. The spectral LDFC range is limited by instrument design considerations (optics, detectors) and stellar leakage brigthness. As the BF intensity increases, detector saturation and response stability requirements become more challenging. To facilitate LDFC implementation, coronagraph masks may be optimized to provide a moderate level of starlight rejection over a wide spectral range outside of the primary science band, enabling efficient use of a wide LDFC spectral range.

\begin{deluxetable*}{lcccccccc}
\tablecaption{LDFC spectral optimization. Stellar absolute magnitutes (top number), wavefront sensing efficiency gain G relative to V band (middle number), and wavefront sensing gain G obtained by changing sensing wavelength from science band to optimal band (bottom number)\label{tab:specopt}}
\startdata
\hline
Band                                    &  U        &   B      &   V      &   R      &   I      &   J      &   H     &   K  \\
$\lambda$ (nm)                          &  366      &  438     &  545     &  641     &  798     & 1220     & 1630    & 2190  \\
Zero Pt ($ph.m^{-2}.s^{-1}.\mu m^{-1}$)    &  7.692e10 & 1.394e11 & 9.962e10 & 7.025e10 & 4.523e10 & 1.933e10 & 9.338e9 & 4.367e9 \\
\hline
Spectral Type &\\
\\
B0V (Teff = 31500 K)                   & -5.374 & -4.307 & -4.000 & -4.000 & -3.645 & -3.270 & -3.111 & -3.044 \\
                                       & {\bf 6.15} & 2.87 & 1.00 & 0.51 & 0.15 & 0.020 & 0.0046 & 0.0011  \\
                                       & 1.00 & 2.14 & 6.15 & 12.06 & 41.00 & 307.5 & 1337.0 & 5591.0  \\
\vspace{-0.2cm}
\\
B5V (Teff = 15700 K)                   & -1.637 & -1.056 & -0.900 & -0.830 & -0.735 & -0.590 & -0.501 & -0.488 \\
                                       & {\bf 3.42} & 2.50 & 1.00 & 0.48 & 0.18 & 0.029 & 0.0073 & 0.0019  \\
                                       & 1.00 & 1.37 & 3.42 & 7.12 & 19.0 & 117.9 & 468.5 & 1000.0  \\
\vspace{-0.2cm}
\\
A0V (Teff =  9700 K)                   & +1.105 & +1.110 & +1.110 & +1.109 & +1.106 & +1.070 & +1.102 & +1.074  \\
                                       & 1.74 & {\bf 2.17} & 1.00 & 0.51 & 0.21 & 0.040 & 0.011 & 0.0028  \\
                                       & 1.25 & 1.00 & 2.17 & 4.25 & 10.2 & 53.9 & 204.7 & 775.0  \\
\vspace{-0.2cm}
\\
A5V (Teff =  8080 K)                   & +2.100 & +2.000 & +1.840 & +1.751 & +1.654 & +1.510 & +1.479 & +1.441  \\
                                       & 1.36 & {\bf 1.87} & 1.00 & 0.55 & 0.25 & 0.052 & 0.015 & 0.0039  \\
                                       & 1.37 & 1.00 & 1.87 & 3.40 & 7.33 & 36.0 & 124.7 & 479.5  \\
\vspace{-0.2cm}
\\
F0V (Teff =  7200 K)                   & +2.857 & +2.804 & +2.510 & +2.344 & +2.171 & +1.900 & +1.802 & +1.757  \\
                                       & 1.26 & {\bf 1.65} & 1.00 & 0.59 & 0.29 & 0.068 & 0.020 & 0.0054  \\
                                       & 1.31 & 1.00 & 1.65 & 2.78 & 5.70 & 24.3 & 82.1 & 305.6  \\
\vspace{-0.2cm}
\\
F5V (Teff =  6510 K)                   & +3.809 & +3.838 & +3.400 & +3.148 & +2.894 & +2.550 & +2.377 & +2.323  \\
                                       & 1.19 & {\bf 1.45} & 1.00 & 0.64 & 0.34 & 0.085 & 0.027 & 0.0073  \\
                                       & 1.22 & 1.00 & 1.45 & 2.26 & 4.30 & 17.1 & 53.9 & 198.6  \\
\vspace{-0.2cm}
\\
G0V (Teff =  5920 K)                   & +5.104 & +5.046 & +4.450 & +4.114 & +3.786 & +3.340 & +3.078 & +3.011  \\
                                       & 0.95 & {\bf 1.25} & 1.00 & 0.69 & 0.39 & 0.11 & 0.037 & 0.010  \\
                                       & 1.32 & 1.00 & 1.25 & 1.80 & 3.20 & 11.6 & 33.7 & 122.5  \\
\vspace{-0.2cm}
\\
G5V (Teff =  5660 K)                   & +5.845 & +5.660 & +4.980 & +4.603 & +4.242 & +3.740 & +3.430 & +3.354  \\
                                       & 0.78 & {\bf 1.16} & 1.00 & 0.72 & 0.42 & 0.12 & 0.044 & 0.012  \\
                                       & 1.49 & 1.00 & 1.16 & 1.61 & 2.78 & 9.56 & 26.5 & 95.9  \\
\vspace{-0.2cm}
\\
K0V (Teff =  5280 K)                   & +7.012 & +6.576 & +5.760 & +5.317 & +4.907 & +4.290 & +3.903 & +3.812  \\
                                       & 0.55 & {\bf 1.02} & 1.00 & 0.77 & 0.47 & 0.15 & 0.058 & 0.016  \\
                                       & 1.85 & 1.02 & 1.02 & 1.33 & 2.20 & 6.80 & 17.6 & 62.6  \\
\vspace{-0.2cm}
\\
K5V (Teff =  4450 K)                   & +9.440 & +8.384 & +7.250 & +6.579 & +6.004 & +5.110 & +4.550 & +4.420  \\
                                       & 0.23 & 0.76 & {\bf 1.00} & 0.95 & 0.67 & 0.28 & 0.13 & 0.037  \\
                                       & 4.35 & 1.32 & 1.00 & 1.05 & 1.49 & 3.57 & 7.94 & 27.2  \\
\vspace{-0.2cm}
\\
M0V (Teff =  3850 K)                   & +11.781 & +10.591 & +9.160 & +8.247 & +7.312 & +6.140 & +5.518 & +5.315 \\
                                       & 0.16 & 0.58 & 1.00 & {\bf 1.18} & 1.16 & 0.63 & 0.30 & 0.094 \\
                                       & 7.61 & 2.03 & 1.18 & 1.00 & 1.02 & 1.89 & 3.93 & 12.6 \\                                       
\vspace{-0.2cm}
\\
M2V (Teff =  3550 K)                   & +12.970 & +11.800 & +10.300 & +9.299 & +8.127 & +6.890 & +6.290 & +6.056 \\
                                       & 0.15 & 0.54 & 1.00 & 1.28 & {\bf 1.57} & 0.90 & 0.42 & 0.14  \\
                                       & 10.5 & 2.88 & 1.57 & 1.22 & 1.00 & 1.74 & 3.74 & 11.6  \\
\vspace{-0.2cm}
\\
M4V (Teff =  3200 K)                   & +15.683 & +14.461 & +12.800 & +11.559 & +9.969 & +8.390 & +7.833 & +7.551   \\
                                       & 0.12 & 0.47 & 1.00 & 1.60 & {\bf 2.87} & 2.25 & 1.02 & 0.34   \\
                                       & 23.5 & 12.5 & 2.87 & 1.80 & 2.87 & 1.28 & 2.83 & 8.41   \\
\vspace{-0.2cm}
\\
M6V (Teff =  2850 K)                   & +19.920 & +18.620 & +16.620 & +14.670 & +12.520 & +10.280 & +9.675 & +9.323   \\
                                       & 0.083 & 0.34 & 1.00 & 3.07 & 9.24 & {\bf 13.3} & 6.28 & 2.25   \\
                                       & 160.2 & 38.7 & 13.3 & 4.33 & 1.44 & 1.00 & 2.12 & 5.91   \\
\vspace{-0.2cm}
\\
M8V (Teff =  2500 K)                   & +21.080 & +21.080 & +18.880 & +16.730 & +14.280 & +11.270 & +10.593 & +10.146   \\
                                       & 0.23 & 0.29 & 1.00 & 3.69 & 14.7 & {\bf 42.8} & 21.6 & 8.46   \\
                                       & 187.4 & 150.0 & 42.8 & 11.6 & 2.92 & 1.00 & 1.98 & 5.07   \\
\enddata
\end{deluxetable*}

\subsection{Spectral Optimization}
Wavefront control performance can be improved by choosing the optimal sensing wavelength independently of the science acquisition spectral band. For example, wavefront sensing sensitivity is especially challenging in the near-IR when observing exoplanets around Sun-like and hotter stars (spectral types G, F, A, B, O), yet the near-IR is spectrally rich in valuable molecular absorption lines. In H-band ($\lambda \approx$ 1.6 $\mu m$), a G0V star yields fewer photons for wavefront sensing than in visible light, and the information content per photon for sensing path length is lower due to the longer wavelength (path length measurement error is proportional to $\lambda / \sqrt{N_{ph}}$).  The underlying assumption behind spectral optimization is that the chromatic relationship between focal plane speckles at different wavelengths is static in time. 

Table \ref{tab:specopt} provides, as a function of stellar spectral type, the relative sensing efficiency for selected wavelength values ranging from 366nm (U band) to 2190nm (K band), assuming a constant $d\lambda/\lambda$ spectral band. For each entry in the table (combination of stellar type and wavelength value), three values are provided:
\begin{itemize}
\item{Top: Absolute stellar magnitude}
\item{Middle: Wavefront sensing efficiency relative to V band}
\item{Bottom: Relative gain in wavefront sensing efficiency obtained by changing sensing wavelength from the science observation wavelength to the optimal sensing wavelength}
\end{itemize}

Flux zero point values are derived from \cite{1998A&A...333..231B} and absolute magnitudes from \cite{2013ApJS..208....9P}. The table shows that for a G0V type star, at equally broad spectral band ($d\lambda/\lambda$), if wavefront sensing were performed at 0.438 $\mu m$, it would be 25\% more efficient than at V band. More importantly, sensing at 1.63 $\mu m$ would be 1/122th as efficient as in B band: a B band wavefront sensor reaches in 1 second the same photon-noise limited measurement precision as a H band sensor in 2 minutes. The wavelength optimization gain is most extreme for near-IR observation of hot stars, where short wavelength sensing provided by LDFC can offer three orders of magnitude boost in efficiency over convential speckle modulation operating at the science acquisition wavelength (775x gain for K band observation of a A0V star by moving wavefront sensing wavelength to B band).

A similar multi-wavelength approach is routinely used on most ground-based adaptive optics systems, which typically sense wavefronts in visible light and perform science acquisition in the near-IR.

\subsubsection{Polarization}
In ultra-high contrast imaging systems, reflection on surfaces at non-normal incidence creates polarization-dependent wavefront errors \citep{2015PASP..127..445B}. A linear polarizer may be required to isolate a single polarization state for high contrast imaging, resulting in a 50\% loss in efficiency. LDFC does not require polarization selection, as wavefront errors due to changes in optics shapes or atmosphere turbulence are independent of polarization.

\subsection{Discussion}

\begin{deluxetable}{lcccc}

\tablecaption{LDFC sensing efficiency gains \label{tab:sensgain}}
\startdata
\hline
                    & SpeLDFC & SpaLDFC & Spe+Spa\\
\hline
\hline
Coh. mix. \& duty c.     & $\approx$ 2x    &   $\approx$ 2x & $\approx$ 2x\\
Spectral Bandwidth  & $\approx$ 4x    &      -      &  $\approx$ 4x\\
Spectral Optim.  & 1-1000x         &   -     & 1-1000x\\
Polarization        & 1-2x            &  1-2x      & 1-2x\\
Spatial gain        & -           &   1-2x   & 1-2x\\
\hline
Total               & $\approx$ 8 - 16,000x &  $\approx$ 2 - 8x & $\approx$ 8-32,000x
\enddata

\end{deluxetable}

Efficiency gains offered by spectral and spatial LDFC are summarized and combined in table \ref{tab:sensgain}. The overall gain ranges from 8x to 32,000x for combined spectral and spatial LDFC, depending on the optical system configuration and target. The largest potential efficiency gain comes from spectral optimization. A narrower $\approx$ 100x to $\approx$ 1000x range wavefront sensing efficiency gain is representative of most systems, where spectral optimization offers a $10-30 \times$ gain and all other gains contribute an additional $10-30 \times$ factor.

We quantify in this section the benefits of this wavefront sensing efficiency gain, noted $G$. We consider the regime where residual wavefront error is dominated by wavefont sensor photon noise and by time lag. We adopt a simplified control loop model, where the wavefront sensing is performed over an exposure time $t$ and the corresponding correction applied with an effective time lag equal to the sensing time $t$. Increasing $t$ reduces photon noise but increases time lag error, so we first estimate the optimal choice for $t$. We consider the case where wavefront aberrations are slower than the sensing rate, so the value of a wavefont mode can be approximated as a linear function of time over $t$, with temporal derivative $a$. We assume the wavefront sensor is photon noise-limited: the measurement noise is $b/\sqrt{t}$ ($b$ is the measurement standard deviation for a 1-sec exposure with a static wavefront error), where $1/b^2$ is proportional to the source flux $F$ and wavefront sensor sensitivity gain $G$:
\begin{equation}
b \propto \sqrt{\frac{1}{G F}}
\end{equation}

The residual wavefront error is the quadratic sum of time lag and photon noise:
\begin{equation}
\sigma^2 = a^2 t^2 + b^2/t
\end{equation}

The optimal sensing time $t$ corresponds to the minimum value of $\sigma$, and is obtained when the derivative of $\sigma^2$ against $t$ is zero: 
\begin{equation}
t = 2^{-1/3} \left( \frac{b}{a} \right)^{2/3} \propto  a^{-2/3} F^{-1/3} G^{-1/3} 
\end{equation}

Omitting the constant factor, we then find 


\begin{equation}
\sigma^2 \propto a^{2/3} b^{4/3} \propto  a^{2/3} F^{-2/3} G^{-2/3}
\end{equation}
 
A gain $G=100$ in sensing sensivity corresponds to the following performance gain in a high contrast imaging system:
\begin{itemize}
\item{The control loop operates $G^{1/3}=4.6$ times faster yielding a $G^{2/3} = 22 \times$ gain in image contrast.}
\item{At equal contrast, wavefront variation timescale is relaxed by a factor $G=100 \times$.}
\item{A given contrast level can be achieved on sources $G=100$ times fainter, or, equivalently, sources $\sqrt{G}$ more distant. In a contrast-limited regime, the corresponding number of sources accessible within the $\sqrt{G}$-radius sphere is multiplied by $G^{3/2} = 1000 \times$.}
\end{itemize}

\section{Practical Motivations for linear image-based sensing}
\label{sec:imsensingbenefits}

\subsection{Control Loop Stability}

LDFC relies entirely on camera calibration for wavefront measurement, while DM modulation schemes such as EFC rely on both DM and camera calibration to yield an accurate estimate of wavefront errors. Detector calibration is usually more accurate and stable than DM calibration, as DMs can experience hysteresis, creep, and temperature-dependence gain, so the detector-based control loop is more likely to be robust. The linear control loop is also simpler than a non-linear EFC control requiring multiple images to be processed to infer wavefront commands. Linear control is a well-understood field with a strong heritage, and can include powerful diagnostics tools as well as performance enhancements such as predictive control.

LDFC is largely immune to detector readout noise and photon noise from incoherent background (zodi/exozodi, dark current) thanks to the bright speckle and the corresponding signal amplification by coherent mixing. While DM modulation sensing schemes can achieve the same insensitvity by using large DM probes, non-ideal DM characteristics (hysteresis, time relaxation) are significant issues when attempting to return to the exact high contrast state following large DM excursions.

\subsection{Image-based DM probing}
\label{ssec:imbasedDMprob}

DM probing is required to drive the imaging system to a high contrast state, and may be necessary to periodically re-calibrate the convergence point of the LDFC loop if the DF-to-BF relationship drifts. DM probing is usually performed as a sequence of DM offsets and image integrations. This process is time-consuming on faint sources, and changes in wavefront state and/or uncalibrated DM drifts during the sensing process can compromise the wavefront estimation. 

Under the LDFC framework, DM probing can be implemented as BF offsets: simultaneously to offsetting the DM, the corresponding BF image offsets are added to the LDFC loop. The BF image offsets can be computed or measured by recording the BF image immediately after the DM command has been issued. This scheme allows for uniterrupted LDFC wavefront stabilization during the DM modulation process, and ensures that the applied wavefront offsets (linear sum of DM probes and incoming wavefront changes) are stable during the measurement process. The approach allows for long integration of the images corresponding to the DM probes (useful when small DM probes are applied), and addresses DM stability concerns by allowing large probes to be used without risks that system cannot return to the high contrast state.

\section{Segmented aperture cophasing for high contrast imaging}
\label{sec:numsimu}

We demonstrate in this section spectral LDFC operation on a simulated coronagraph for a segmented aperture. We consider a 12m diameter segmented aperture observing a $m_V = 5$ star with a 40\% efficiency (excluding losses due to coronagraph masks). The Adaptive Optics Control Computation Engine \citep{AOCCE} software package was used to perform LDFC simulations.

\begin{figure}[htb]
\includegraphics[scale=0.34]{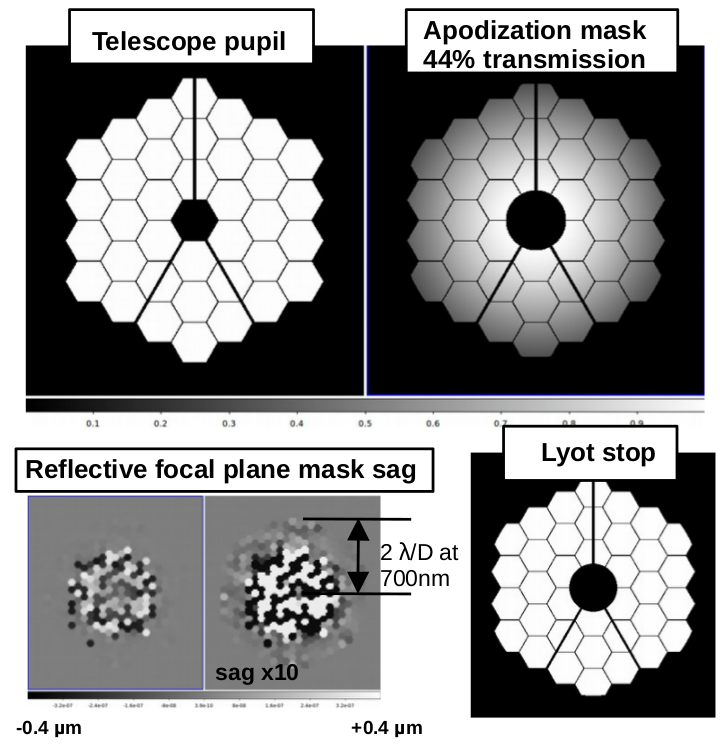} 
\caption[LDFCprincip] { \label{fig:coro} Apodized Pupil Lyot Complex mask coronagraph (APLCMC) design. The input segmented pupil (top left) is first amplitude-apodized (top right). A computer-optimized multi-zone reflective focal plane mask (bottom left) diffracts starlight out of the Lyot stop transmission area (bottom right).}
\end{figure}

\subsection{Starlight Suppression System}

We adopt a APLCMC-type coronagraph \citep{2014ApJ...780..171G}, consisting of a pupil plane apodizer, a focal plane mask, and a Lyot stop. The input pupil consists of 36 hexagonal segments, representative of a future large space-based telescope \citep{2016SPIE.9904E..4RC}. A circular central obstruction is held by three spiders vanes. Coronagraph masks are shown in Fig. \ref{fig:coro}. The apodization mask has a 44\% transmission and contains no high spatial frequency features. The focal plane mask is realized as a reflective mirror consisting of several hundred hexagonal zones, each with a different height. The hexagon heights have been computer-optimized to produce a deep null in the geometric pupil across a 10\% wide spectral band by destructive interfence of ligh incident on the hexagons. As the mask does not absorb light, most starlight is diffracted out of the geometric pupil and blocked by the Lyot stop.

\begin{figure}[htb]
\includegraphics[scale=0.28]{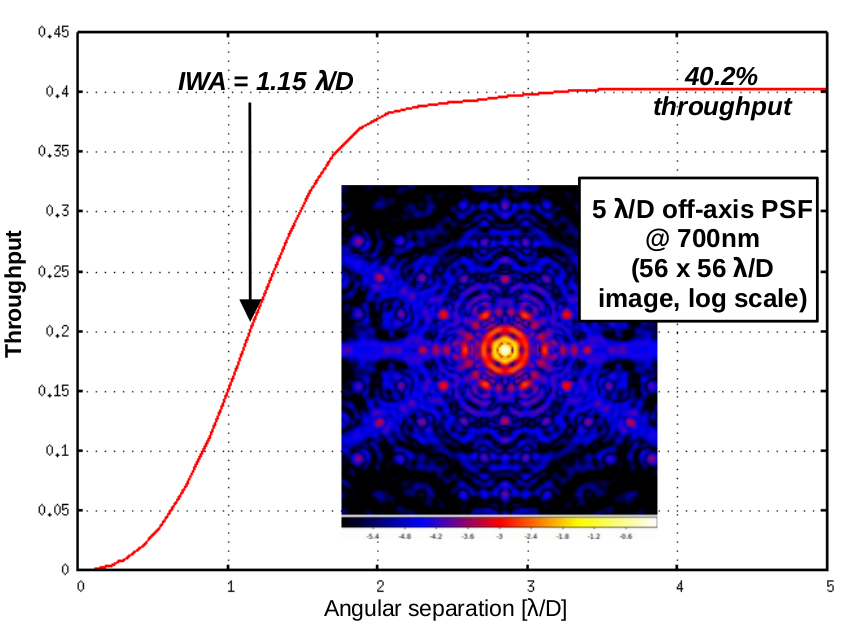} 
\caption[LDFCprincip] { \label{fig:coroT} Coronagraph throughput and off-axis PSF (log scale).}
\end{figure} 

The coronagraph throughput is defined here as the total fraction of light from a point source reaching the final focal plane. At large angular separation, it is limited to $\approx$ 40 \% (Fig. \ref{fig:coroT}) due to absorption by the pupil apodization mask, as well as some loss at the Lyot stop. The inner working angle (IWA), where the throughput is half of this level, is 1.15 $\lambda/D$. Thanks to the limited extent of the focal plane mask, beyond $\approx$ 3 $\lambda/D$ separation, the off-axis PSF (shown in Fig. \ref{fig:coroT} at a 5 $\lambda/D$ separation) is largely translation invariant, and the throughput is nearly constant.

The raw PSF contrast is $\approx$ 1e-10 in the 10\% wide band for which the focal plane mask is optimized, and residual starlight intensity gradually increases at wavelengths farther from the high contrast band. The top part of Fig. \ref{fig:LDFCcalib} shows stellar residual light through the coronagraph across a 30\% wide band centered on the high contrast band, with wavelength increasing in 1\% increments from left to right and then from bottom to top, starting at 598nm (lower left image) up to 801nm (top right). The images represent data collected by an integral field spectrograph at spectral resolution $R = 100$.

\subsection{Calibration}

\begin{figure*}[htb]
\includegraphics[scale=0.34]{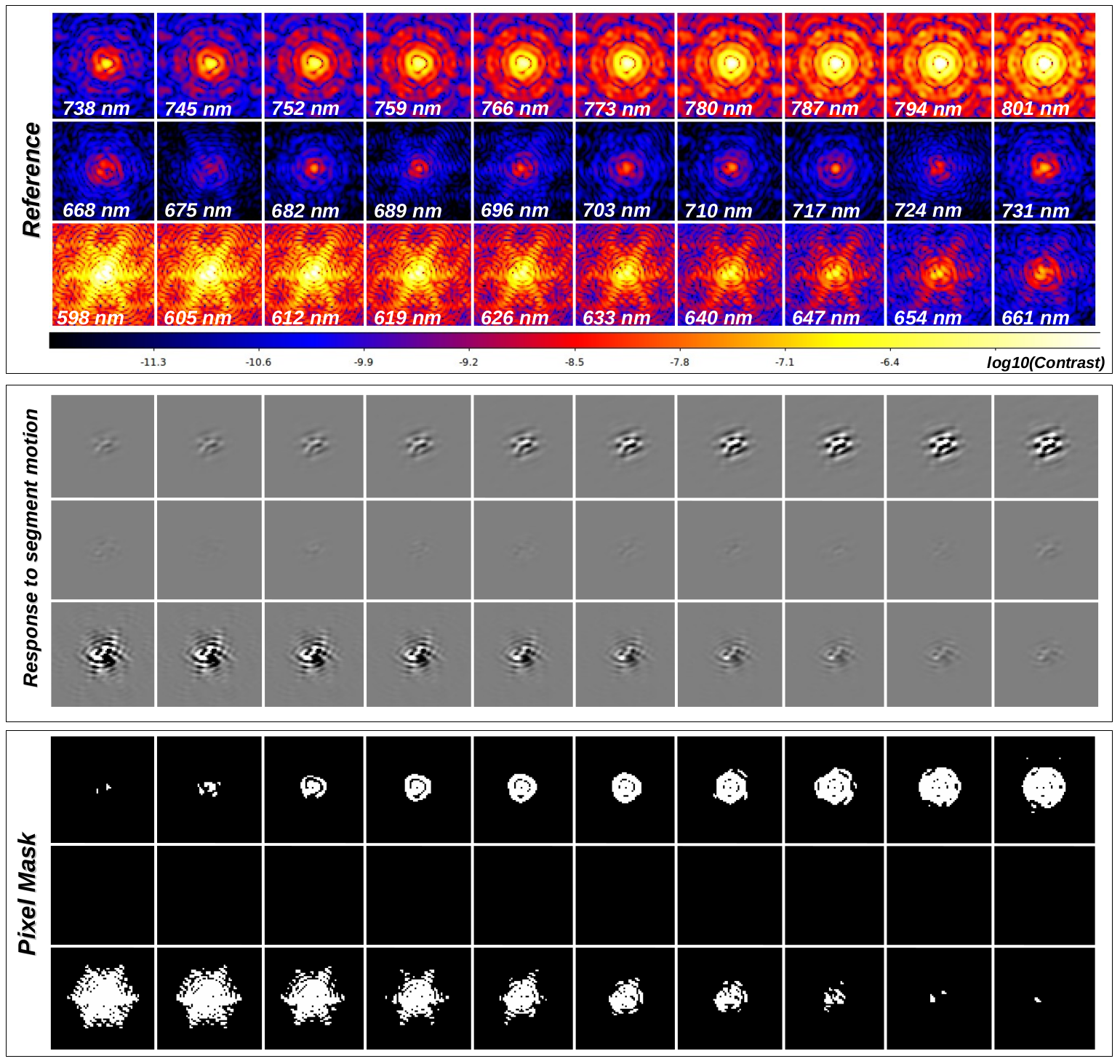} 
\caption[LDFCprincip] { \label{fig:LDFCcalib} Spectral Linear Dark Field Control principle calibration. Top: Coronagraphic image of an on-axis point source, used as a reference for the control loop. Thirty images are shown, each corresponding to a wavelength. The coronagraph is optimized to deliver high contrast over the central 10\% spectral band. Starlight leakage increases as other wavelengths. Middle: Linear response to a single wavefront mode. 111 such responses are computed or measured for calibration. Bottom: WFS pixel mask, showing which pixels are taken into acount for wavefront measurement.}
\end{figure*}

LDFC calibration consists of the reference detector intensity image and the linear response to each wavefront mode to be controlled. Simulated images acquired simultaneously at 30 wavelengths from 598nm to 801nm with a flat wavefront are shown in Fig. \ref{fig:LDFCcalib}, top. We only consider here segment cophasing and tip-tilt errors, so all wavefront errors are represented as a linear combination of $37 \times 3 = 111$ modes. We note that the central segment is not illuminated, and that overall piston has no effect on the image, so we expect 107 modes to represent all possible wavefront realizations.

For each segment motion, the corresponding BF response is measured as the difference between the PSF set with aberration and the reference PSF set. One of 111 such linear responses is shown in Fig. \ref{fig:LDFCcalib} (center): the 30 images in this middle panel are the PSF response to the same mode, but at 30 different wavelength values. Finally, a binary mask is constructed to select image pixels that are used for sensing. The mask, shown in Fig. \ref{fig:LDFCcalib} (bottom), is common to all modes, and is constructed by selecting pixels that have the strongest linear response to segment motions. The mask is computed by thresholding a map of linear response amplitude squared, summed over all 111 mirror modes. The central high contrast 10\% bandwidth does not participate to the measurement, and most of the signal originates from the outer edges of the 30\% wide spectral band, where coronagraph leaks are strongest.

\subsection{Closed Loop Operation}

Closed loop control follows steps 6 and 7 listed in section \ref{ssec:algo}. Images, each consisting of 30 wavelength slices, are acquired at regular intervals. Each image is reference-subtracted. The values over active pixels (pixels that are "on" in the pixel mask) are arranged as a 1D vector, and multiplied by the control matrix (obtained by pseudo-inverse of the calibration response matrix). The resulting control command then moves the segments in tip, tilt and piston.

\begin{figure*}[htb]
\includegraphics[scale=0.3]{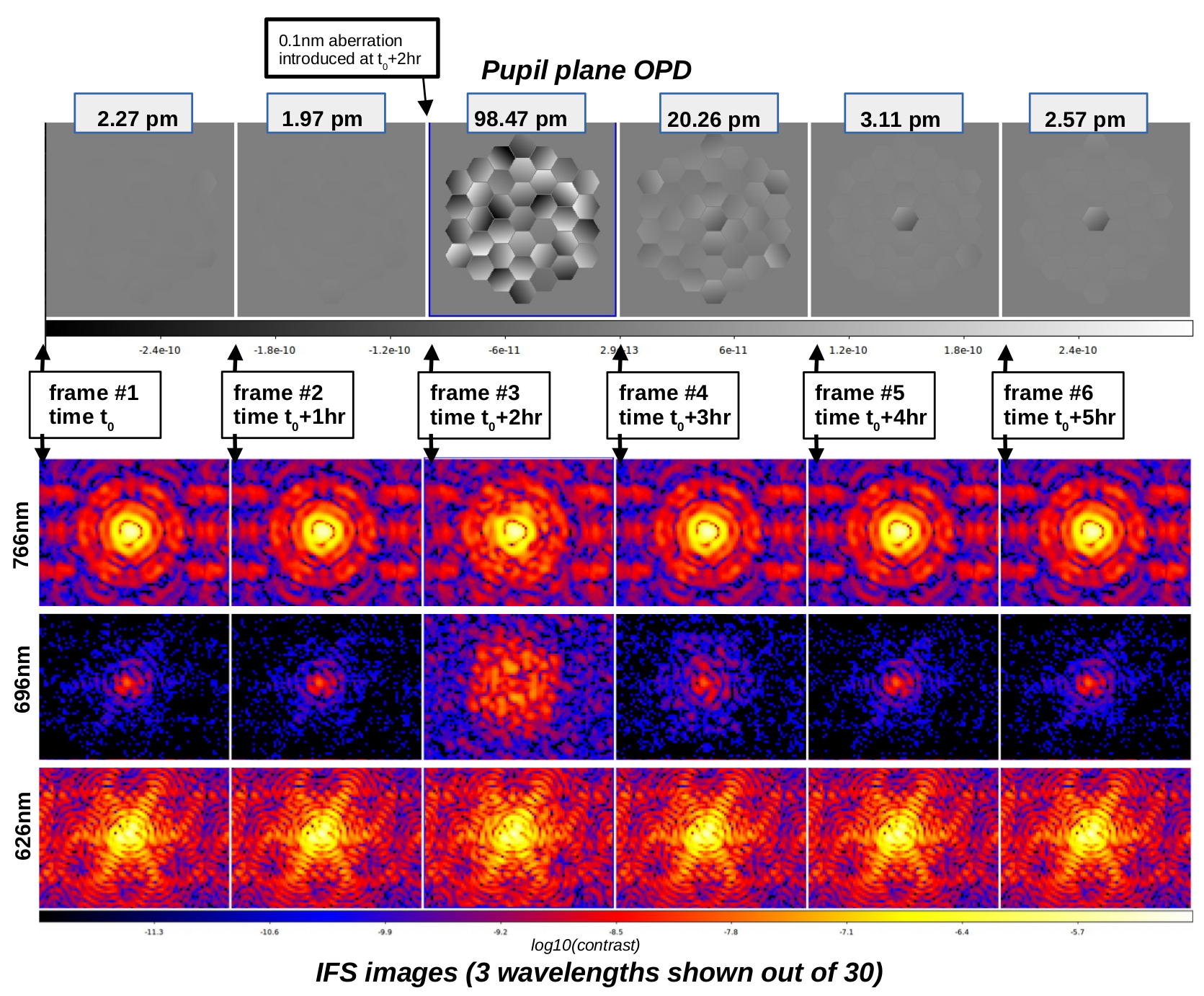} 
\caption[LDFCprincip] { \label{fig:pupfocseq} Closed loop sequence, showing pupil plane optical path length difference (top) and selected wavelength images (bottom). A 0.1nm disturbance in introduced at the end of the second 1 hr exposure, and subsequently corrected by LDFC. Note that only 3 wavelenths are shown, yet all 20 wavelenths participate to the wavefront measurement (see pixel mask on Fig. \ref{fig:LDFCcalib}). LDFC operates here on 1-hr long exposures, and the corresponding correction is applied at the end of the measurement exposure before the next 1-hr exposure stars. Time between the end of an exposure and the corresponding correction is assumed here to be negligible.}
\end{figure*}

Figure \ref{fig:pupfocseq} shows a 6hr-long simulation sequence, during which a random perturbation was introduced at $t=t_0 + 2hr$. A 12-m diameter aperture observing a $m_V = 5$ source is simulated here, assuming a 40\% efficiency. Each measurement/correction cycle spans one hour. The 0.1nm RMS perturbation, consisting of segment tip-tilt and piston, is corrected in $\approx$ two measurement/correction cycles. Prior to the perturbation, the wavefront error is at $\approx$ 2pm RMS due to propagation of measurement photon noise in the ongoing wavefront control loop. 

The image raw contrast averaged between 2 and 5 $\lambda/D$ radius and within the 10\% central spectral band rises from $\approx$ 1e-10 prior to aberration injection to $\approx$ 1e-8 when the disturbance is introduced. The central mirror segment is left uncorrected, as it is not illuminated (hidden behind the central obstruction).

\begin{figure*}[htb]
\includegraphics[scale=0.25]{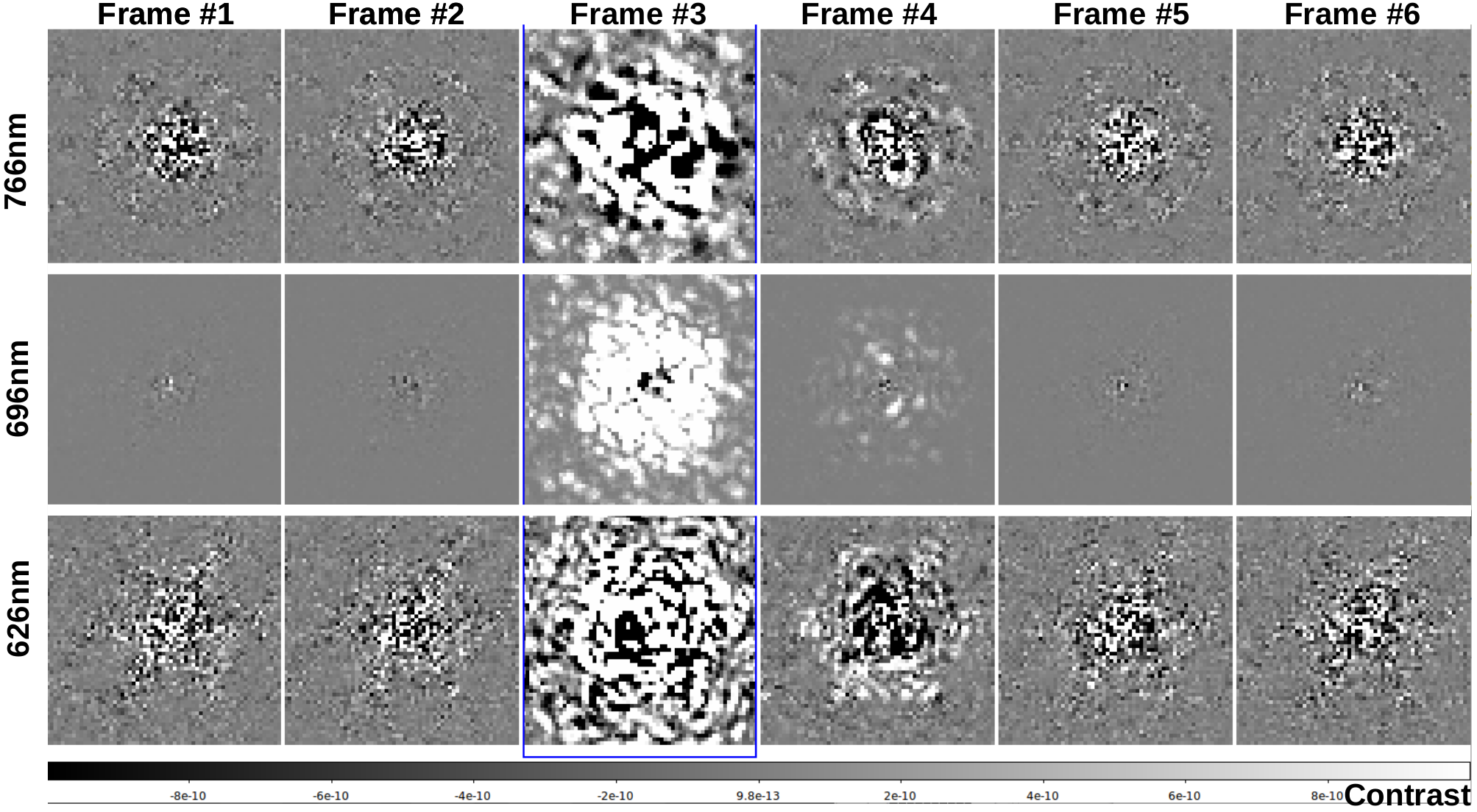} 
\caption[LDFCprincip] { \label{fig:psfsnoise} Reference-subtracted PSF sequence. Only 3 out of the 30 wavelength channels are shown here: 626nm (bottom), 696nm (center) and 766nm (top). The 696nm channel is within the 10\% high contrast spectral band, while the two other channels are brightly illuminated. The wavefront disturbance is injected between frames 2 and 3. Photon noise dominates the residual in frames 1,2,5 and 6, while frames 3 and 4 residuals are dominated by wavefront errors.}
\end{figure*} 

The PSF stability during the sequence is shown in Fig. \ref{fig:psfsnoise}, where a PSF reference has been subtracted. The reference for this subtraction is a 4-hr integration measured as the average between the two PSFs preceeding and the two PSFs following the six PSFs shown in Figs \ref{fig:pupfocseq} and \ref{fig:psfsnoise} (frames -1, 0, 7 and 8). This choice is representative of current PSF calibration and subtraction techniques such as angular differential imaging. High contrast PSFs in frames 1, 2, 5 and 6 are dominated by photon noise due to the finite number of photon in each 1-hr long integration: 2 hrs after the disturbance is introduced, image contrast is no longer limited by wavefront errors.

\begin{figure}[htb]
\includegraphics[scale=0.15]{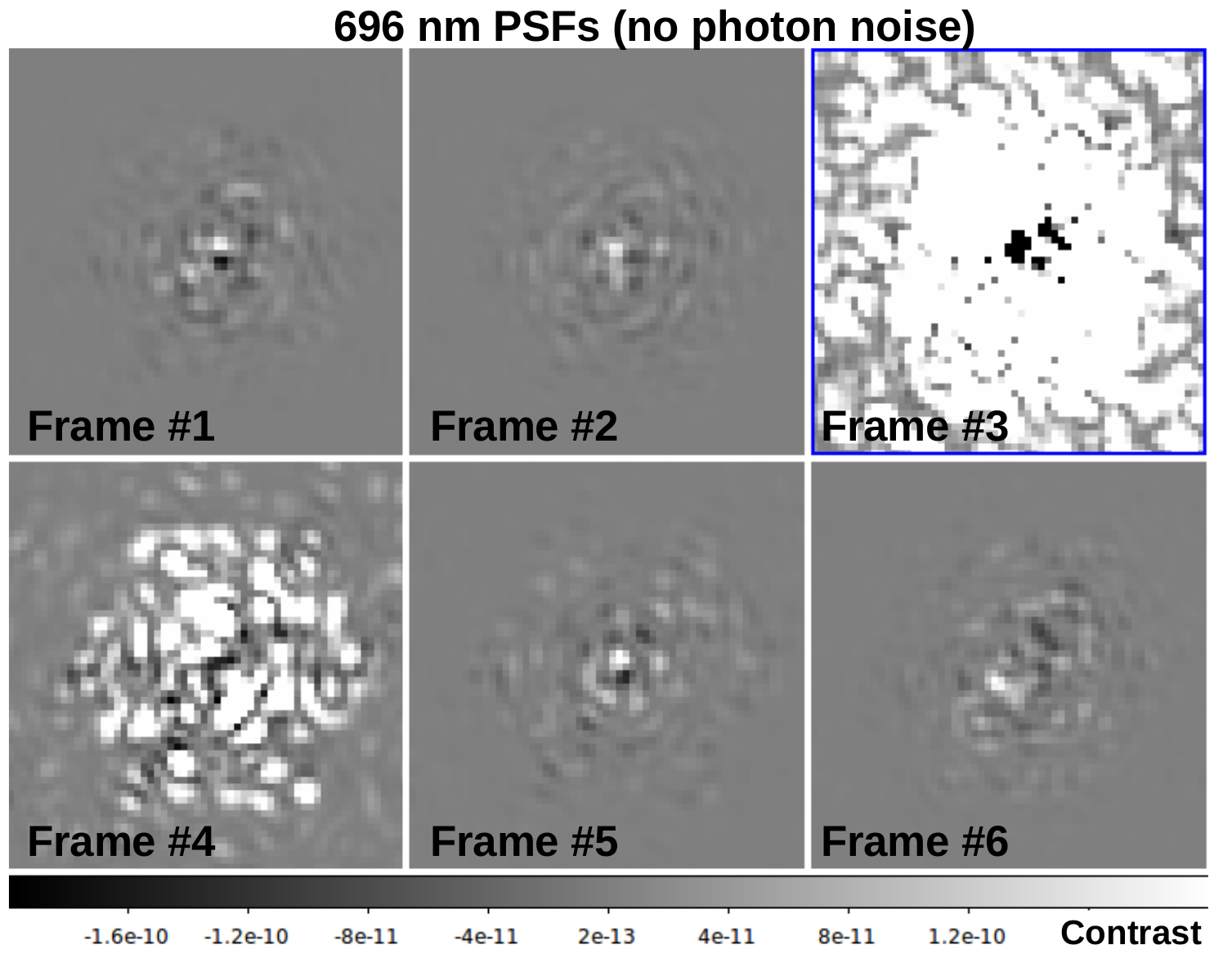} 
\caption[LDFCprincip] { \label{fig:psfsnonoise} Noiseless reference-subtracted PSF sequence in one of the high contrast wavelengths. This figure shows the middle row of Fig. \ref{fig:psfsnoise} with photon noise removed from the images.}
\end{figure} 

\begin{figure*}[htb]
\includegraphics[scale=0.196]{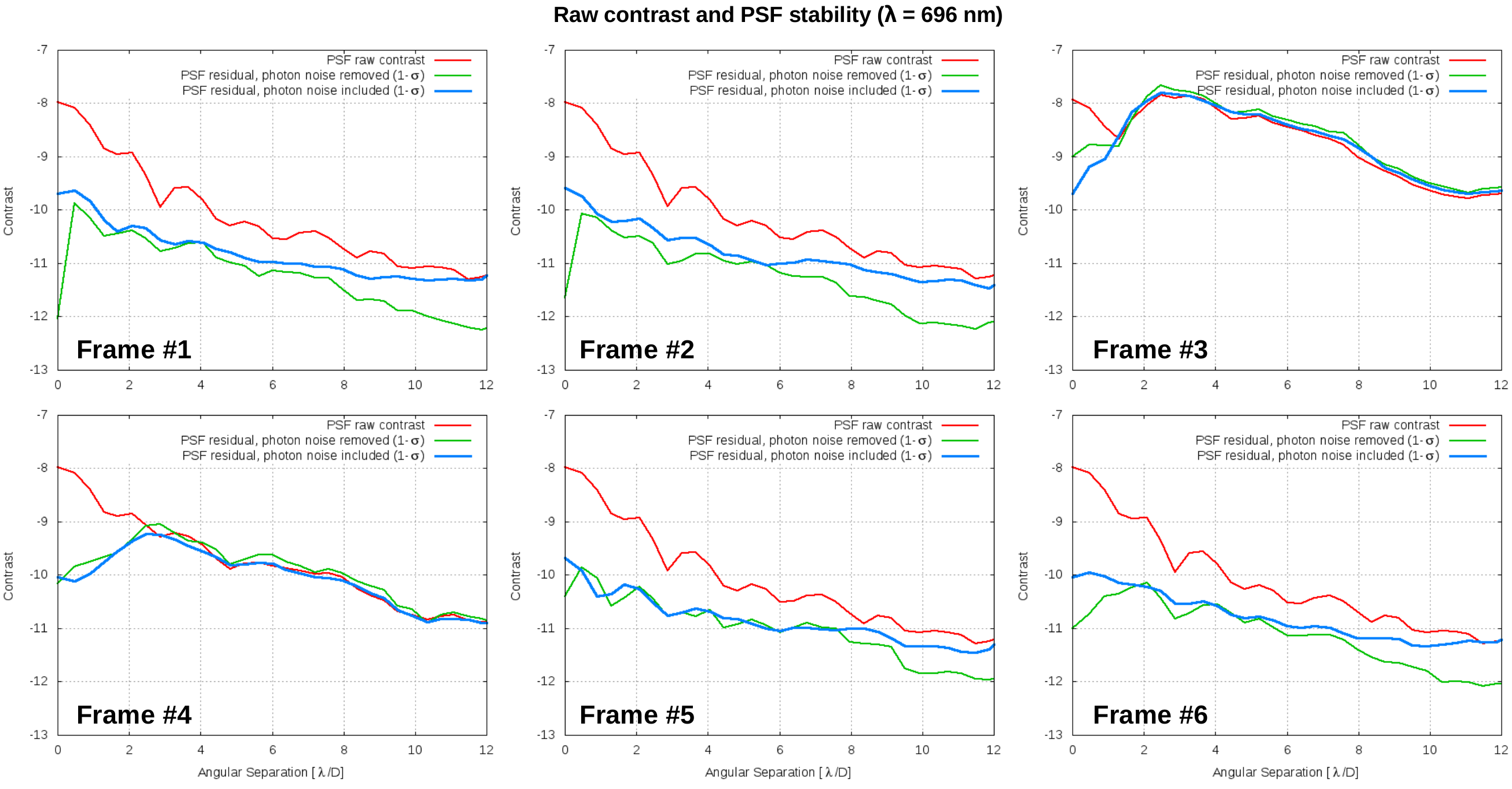} 
\caption[PSFcontrast] { \label{fig:PSFcontrast} Raw (red) and residual (blue and green) contrast profiles for one of the high contrast wavelength channels. The green curve shows the residual contrast profile (standard deviation along a circle centered on the optical axis) in the absence of image photon noise. The blue curve includes image photon noise. Both curves do include the effect of photon noise in the LDFC wavefront sensing.}
\end{figure*}

Noiseless high contrast reference-subtracted PSFs, shown in Fig. \ref{fig:psfsnonoise}, reveal the contrast floor due to wavefront error below the photon noise limit. Corresponding contrast curves for one of the high contrast wavelengths (696nm) are shown in Fig. \ref{fig:PSFcontrast}. In the absence of aberrations (frames 1 and 2), the noise-free PSF contrast is $\approx \pm $ 4e-11 in each 1-hr long exposure, and is largely uncorrelated from frame to frame. This noise floor is due to photon noise in the LDFC sensing, so it is not correlated over multiple cycles. Following frames 2 (uncorrected wavefront disturbance) and 3 (partially corrected), frames 5 and 6 show a similar contrast level.

\subsection{Sensing efficiency}

We can estimate the sensing efficiency and compare it to the fundamental photon noise limit identified in \cite{2005ApJ...629..592G}: with $N_{ph}$ available, a phase and amplitude sensor can ideally measure a single wavefront mode with a $\sqrt{2}/\sqrt{N_{ph}}$ standard deviation.

With a $\approx$ 80 $m^2$ aperture and a 16\% overall efficiency (40\% efficiency and 40\% coronagraph throughput), the $m_R=5$ source provides 8.99e9 $ph.s^{-1}.\mu m^{-1}$ at 700nm, or $N_{ph} = 6.47e12$ for each 1 hr exposure across the two 10\% sensing spectral bands (20\% total). The corresponding expected ideal photon-noise limited measurement noise is $\sqrt{2}/\sqrt{N_{ph}} = 5.6e-7 rad$ RMS, or 0.061 pm RMS per mode assuming a 700nm central wavelength. With 107 modes sensed and corrected, the corresponding overall wavefront measurement noise should ideally be 0.64 pm, while the measured value is $\approx 3.5 \times$ larger.

Spectral LDFC is among the most sensitive WFS options available, as most WFS implementations are far below the ideal theoretical limit \citep{2005ApJ...629..592G}. LDFC's sensitivity may be further improved by adopting a more inclusive pixel selection mask: a with a larger number of "on" pixels (in bottom panel of Fig. \ref{fig:LDFCcalib}), LDFC uses a larger fraction of total starlight.

\section{Limitations and Concerns}
\label{sec:disc}

\subsection{LDFC scope}

LDFC is a wavefront stabilization technique, and (an)other approach(es) must be employed to reach a deep contrast state.  Once deep contrast is obtained, the LDFC reference image should be acquired, and any further LDFC operation will drive the starlight suppression system to this reference. As described in \S\ref{ssec:imbasedDMprob}, LDFC can also assist DM probing schemes aimed at reaching deep contrast: by referencing DM probes to BF images, probing can be made more stable and LDFC can operate during the DM probing sequence. 

LDFC can be extended to multiple sensors to reduce measurement null space and improve sensitivity. The measurement vector, which is assumed in this paper to only consist of post-corongraph BF integral field spectrograph images, can be grown to include other signals linearly coupled to wavefront errors such as light from a coronagraphic low-order wavefront sensor or light from a separate post-coronagraph camera.

\subsection{Reference stability}
\label{ssec:refstab}

LDFC assumes that if the BF intensity is kept stable, then the DF complex amplitude will also be frozen. Combinations of changes in optical alignment, hardware component response variations (for example: camera flat field response, aging of coatings on optical elements), and source characteristics (stellar spectral type) could modify this relationship and slowly drive the LDFC loop away from the ideal DF contrast even if the BF intensity is kept constant. If different cameras are used for LDFC BF signal and science DF acquisition, variations in non-common path optical aberrations could contribute to this issue and must be kept small.

Changes in the LDFC reference will require the deep contrast to be re-established periodically using for example DM probing. The timescale over which the LDFC reference drifts is likely much longer than the wavefront stability timescale, so it is expected that the re-calibration of the reference to a deep-contrast state occurs infrequently compared to LDFC iterations. Quantitative analysis of the high contrast imaging system will be required to define both timescales.

\subsection{Linear response stability}
\label{ssec:respstab}

LDFC assumes stability of the linear response matrix linking BF intensity changes to DF complex amplitude changes. Changes in this response can drive the LDFC loop to poor temporal performance and instability. The timescale over which the linear response changes is likely much longer than both the LDFC iterations and the timescale over which the reference changes (\S\ref{ssec:refstab}). For a closed loop control, response matrix changes at the 10\% level are likely acceptable and will not significantly affect the loop convergence time, although the exact allowable change depends on error propagation/amplification through the response matrix inversion. The most likely source of RM change may be due to properties of the astrophysical source (spectral type, angular size). Full LDFC re-calibration, as described in \S\ref{ssec:algo}, is required to track RM changes.

This issue is not unique to LDFC: all wavefront control schemes rely on stability of the relationship between input measurements and corresponding control signals. For the DM probing approach, this relationship is represented as a response matrix between DM commands and focal plane complex amplitude, and is also assumed to be stable during long periods of time (this matrix is noted the "G-matrix" in \cite{2009SPIE.7440E..0DG}).

\subsection{Measurement null space}
\label{ssec:nullspace}

The measurement null space consists of all wavefront state variations that can change the DF without affecting the BF. In a space telescope system where aberrations can only originate from a few optical components, the measurement null space is likely small with spatial + spectral LDFC, but it must be mapped. Wavefront variations in the null space are not corrected by LDFC, potentially affecting imaging performance.

While the numerical simulation presented in this paper did not reveal the existence of a null space, actual optical systems are considerably more complex. Spatial LDFC is prone to null space issues: in the half-field dark hole configuration, a well-chosen combination of amplitude and phase aberrations can induce a single-sided speckle in the dark hole with little modulation of the BF. Spectral LDFC is more immune to null space issues, as it is difficult for aberrations to only affect a $\approx$ 10\% wide spectral band without affecting other wavelengths. More detailed analysis is required to map null space on realistic optical systems. Combining spectral LDFC with spatial LDFC and/or other linear sensors should further reduce measurement null space.

\subsection{Acquiring LDFC signal: practical challenges}
Data acquisition must overcome the BF/DF flux contrast and use optical/detector hardware with minimum impact on science operation and little non-common path errors. For example, the flux ratio between in-band and out-of-band images in Fig. \ref{fig:LDFCcalib} (spectral LDFC) is approximately 3 orders of magnitude, and may exceed the dynamical range of an IFS. Some of the bright light may contaminate the high contrast spectral band due to optical crosstalk; readout cadence for photon counting in the high contrast area will be too slow for the bright areas; detector traps could cause out-of-band light to affect the in-band faint signal. A separate camera or IFS may therefore be required to capture the out-of-band light, or an absorption filter could be deployed to bring the BF within the camera's tolerance.

\subsection{Acquiring or computing the calibration}
Acquiring the reference frame and intensity derivatives may be challenging in a real imaging system due to measurement noise and limited dynamical range. While signal averaging can overcome most sources of noise, the required calibration acquisition time may become prohibitive. Hardware limitations such as DM hysteresis and limited DM speed may also contribute to this issue. 

In a high contrast imaging system, both the reference and linear response can be computed from the coronagraph design, as the BF component is due to well-understood limitations of the optical system, as opposed to the DF light, which is due to poorly understood wavefront errors.

\section{Conclusion and future work}

\subsection{LDFC operation in high contrast imaging systems}
LDFC can provide a significant boost in wavefront sensing sensitivity and control stability. While further system-specific detailed analysis is required to address and quantify the main outstanding issues (practical considerations, measurement null space), the technique is a powerful addition to existing DM probing approaches.

LDFC, as a differential loop for contrast stabilization, must be used alongside an absolute control loop able to drive the starlight suppression system to a high contrast state. The two approaches offer complementary benefits: LDFC provides efficiency, sensitivity and fast temporal bandwidth while the absolute control loop can track slow changes in the BF-DF relationship. Both loops can run simultaneously as described in \S\ref{ssec:imbasedDMprob}.
Ideally, the time-consuming absolute loop can be run on a bright star, and the observatory can then rely entirely on the relative loop for fainter targets.

\subsection{PSF calibration}

As future work, it would be useful to explore how BF signals can be used to estimate the intensity map in the DF, so that it can be numerically subtracted from the actual images to increase the contrast detection limit. The DF complex amplitude can be derived from BF intensity through a linear relationship, and can then be squared (to yield intensity) and subtracted from the high contrast science images. The required linear calibration between BF intensity and DF complex amplitude can be either computed using a model of the coronagraph system or acquired by modulating the DM(s) actuators. PSF calibration is a powerful extension of the LDFC approach as it mitigates temporal control bandwidth limitations:
\begin{itemize}
\item{Wavefront variations occurring on a timescale comparable to the LDFC signal acquisition are too fast to be corrected (and could even be amplified by resonance with the control loop), but are still measured. Their contribution to the PSF can be subtracted.}
\item{ The starlight suppression system may experience aberrations that cannot be corrected by DM(s), yet have a measurable effect of the BF. Provided that a calibration exists between these modes and the DF intensity or complex amplitude, their contribution to the PSF can be removed.}
\end{itemize}
High contrast imaging PSF calibration from linear telemetry has been demonstrated by \cite{2011PASP..123.1434V} using an empirical algorithm that does not require knowledge of the linear relationship between BF intensity and DF complex amplitude. The same approach can be applied here.

\acknowledgments    

Linear algebra computations for this work were performed with GPU acceleration provided by the MAGMA library\citep{tdb10,tnld10,dghklty14,ntd10,ntd10_vecpar}. This material is based upon work supported by NASA's Exoplanet Exploration program under the Segmented Coronagraph Design and Analysis task. The  authors  acknowledge  support  from  the  Japan Society for the Promotion of Science (JSPS, Grant-in-Aid for Research 23340051 and 26220704).



\bibliography{ms}

\begin{thebibliography}{26}
\expandafter\ifx\csname natexlab\endcsname\relax\def\natexlab#1{#1}\fi

\bibitem[{{Bessell} {et~al.}(1998){Bessell}, {Castelli}, \&
  {Plez}}]{1998A&A...333..231B}
{Bessell}, M.~S., {Castelli}, F., \& {Plez}, B. 1998, \aap, 333, 231

\bibitem[{{Bord{\'e}} \& {Traub}(2006)}]{2006ApJ...638..488B}
{Bord{\'e}}, P.~J. \& {Traub}, W.~A. 2006, \apj, 638, 488

\bibitem[{{Bottom} {et~al.}(2017){Bottom}, {Wallace}, {Bartos}, {Shelton}, \&
  {Serabyn}}]{2017MNRAS.464.2937B}
{Bottom}, M., {Wallace}, J.~K., {Bartos}, R.~D., {Shelton}, J.~C., \&
  {Serabyn}, E. 2017, \mnras, 464, 2937

\bibitem[{{Breckinridge} {et~al.}(2015){Breckinridge}, {Lam}, \&
  {Chipman}}]{2015PASP..127..445B}
{Breckinridge}, J.~B., {Lam}, W.~S.~T., \& {Chipman}, R.~A. 2015, \pasp, 127,
  445

\bibitem[{{Crooke} {et~al.}(2016){Crooke}, {Roberge}, {Domagal-Goldman},
  {Mandell}, {Bolcar}, {Rioux}, {Perez}, \& {Smith}}]{2016SPIE.9904E..4RC}
{Crooke}, J.~A., {Roberge}, A., {Domagal-Goldman}, S.~D., {Mandell}, A.~M.,
  {Bolcar}, M.~R., {Rioux}, N.~M., {Perez}, M.~R., \& {Smith}, E.~C. 2016, in
  \procspie, Vol. 9904, Space Telescopes and Instrumentation 2016: Optical,
  Infrared, and Millimeter Wave, 99044R

\bibitem[{Dongarra {et~al.}(2014)Dongarra, Gates, Haidar, Kurzak, Luszczek,
  Tomov, \& Yamazaki}]{dghklty14}
Dongarra, J., Gates, M., Haidar, A., Kurzak, J., Luszczek, P., Tomov, S., \&
  Yamazaki, I. 2014, Numerical Computations with GPUs, 1

\bibitem[{{Give'On}(2009)}]{2009SPIE.7440E..0DG}
{Give'On}, A. 2009, in \procspie, Vol. 7440, Techniques and Instrumentation for
  Detection of Exoplanets IV, 74400D

\bibitem[{{Guyon}(2005)}]{2005ApJ...629..592G}
{Guyon}, O. 2005, \apj, 629, 592

\bibitem[{{Guyon}(2017)}]{AOCCE}
---. 2017, Adaptive Optics Control Computational Engine (AOCCE),
  \url{https://github.com/oguyon/AdaptiveOpticsControl}

\bibitem[{{Guyon} {et~al.}(2014){Guyon}, {Hinz}, {Cady}, {Belikov}, \&
  {Martinache}}]{2014ApJ...780..171G}
{Guyon}, O., {Hinz}, P.~M., {Cady}, E., {Belikov}, R., \& {Martinache}, F.
  2014, \apj, 780, 171

\bibitem[{{Guyon} {et~al.}(2009){Guyon}, {Matsuo}, \&
  {Angel}}]{2009ApJ...693...75G}
{Guyon}, O., {Matsuo}, T., \& {Angel}, R. 2009, \apj, 693, 75

\bibitem[{{Levine} {et~al.}(2009){Levine}, {Lisman}, {Shaklan}, {Kasting},
  {Traub}, {Alexander}, {Angel}, {Blaurock}, {Brown}, {Brown}, {Burrows},
  {Clampin}, {Cohen}, {Content}, {Dewell}, {Dumont}, {Egerman}, {Ferguson},
  {Ford}, {Greene}, {Guyon}, {Hammel}, {Heap}, {Ho}, {Horner}, {Hunyadi},
  {Irish}, {Jackson}, {Kasdin}, {Kissil}, {Krim}, {Kuchner}, {Kwack}, {Lillie},
  {Lin}, {Liu}, {Marchen}, {Marley}, {Meadows}, {Mosier}, {Mouroulis},
  {Noecker}, {Ohl}, {Oppenheimer}, {Pitman}, {Ridgway}, {Sabatke}, {Seager},
  {Shao}, {Smith}, {Soummer}, {Stapelfeldt}, {Tenerell}, {Trauger}, \&
  {Vanderbei}}]{2009arXiv0911.3200L}
{Levine}, M., {Lisman}, D., {Shaklan}, S., {Kasting}, J., {Traub}, W.,
  {Alexander}, J., {Angel}, R., {Blaurock}, C., {Brown}, M., {Brown}, R.,
  {Burrows}, C., {Clampin}, M., {Cohen}, E., {Content}, D., {Dewell}, L.,
  {Dumont}, P., {Egerman}, R., {Ferguson}, H., {Ford}, V., {Greene}, J.,
  {Guyon}, O., {Hammel}, H., {Heap}, S., {Ho}, T., {Horner}, S., {Hunyadi}, S.,
  {Irish}, S., {Jackson}, C., {Kasdin}, J., {Kissil}, A., {Krim}, M.,
  {Kuchner}, M., {Kwack}, E., {Lillie}, C., {Lin}, D., {Liu}, A., {Marchen},
  L., {Marley}, M., {Meadows}, V., {Mosier}, G., {Mouroulis}, P., {Noecker},
  M., {Ohl}, R., {Oppenheimer}, B., {Pitman}, J., {Ridgway}, S., {Sabatke}, E.,
  {Seager}, S., {Shao}, M., {Smith}, A., {Soummer}, R., {Stapelfeldt}, K.,
  {Tenerell}, D., {Trauger}, J., \& {Vanderbei}, R. 2009, ArXiv e-prints

\bibitem[{{Malbet} {et~al.}(1995){Malbet}, {Yu}, \&
  {Shao}}]{1995PASP..107..386M}
{Malbet}, F., {Yu}, J.~W., \& {Shao}, M. 1995, \pasp, 107, 386

\bibitem[{{Martinache} {et~al.}(2016){Martinache}, {Jovanovic}, \&
  {Guyon}}]{2016A&A...593A..33M}
{Martinache}, F., {Jovanovic}, N., \& {Guyon}, O. 2016, \aap, 593, A33

\bibitem[{{Miller} {et~al.}(2017){Miller}, {Guyon}, \&
  {Males}}]{2017arXiv170304259M}
{Miller}, K., {Guyon}, O., \& {Males}, J.~R. 2017, ArXiv e-prints

\bibitem[{Nath {et~al.}(2010{\natexlab{a}})Nath, Tomov, \&
  Dongarra}]{ntd10_vecpar}
Nath, R., Tomov, S., \& Dongarra, J. 2010{\natexlab{a}}, in Proceedings of the
  2009 International Meeting on High Performance Computing for Computational
  Science, VECPAR'10 (Berkeley, CA: Springer)

\bibitem[{Nath {et~al.}(2010{\natexlab{b}})Nath, Tomov, \& Dongarra}]{ntd10}
Nath, R., Tomov, S., \& Dongarra, J. 2010{\natexlab{b}}, Int. J. High Perform.
  Comput. Appl., 24, 511

\bibitem[{{Pecaut} \& {Mamajek}(2013)}]{2013ApJS..208....9P}
{Pecaut}, M.~J. \& {Mamajek}, E.~E. 2013, \apjs, 208, 9

\bibitem[{{Shaklan} {et~al.}(2005){Shaklan}, {Marchen}, {Green}, \&
  {Lay}}]{2005SPIE.5905..110S}
{Shaklan}, S.~B., {Marchen}, L., {Green}, J.~J., \& {Lay}, O.~P. 2005, in
  \procspie, Vol. 5905, Techniques and Instrumentation for Detection of
  Exoplanets II, ed. D.~R. {Coulter}, 110--121

\bibitem[{{Shi} {et~al.}(2016){Shi}, {Balasubramanian}, {Hein}, {Lam}, {Moore},
  {Moore}, {Patterson}, {Poberezhskiy}, {Shields}, {Sidick}, {Tang}, {Truong},
  {Wallace}, {Wang}, \& {Wilson}}]{2016JATIS...2a1021S}
{Shi}, F., {Balasubramanian}, K., {Hein}, R., {Lam}, R., {Moore}, D., {Moore},
  J., {Patterson}, K., {Poberezhskiy}, I., {Shields}, J., {Sidick}, E., {Tang},
  H., {Truong}, T., {Wallace}, J.~K., {Wang}, X., \& {Wilson}, D. 2016, Journal
  of Astronomical Telescopes, Instruments, and Systems, 2, 011021

\bibitem[{{Singh} {et~al.}(2015){Singh}, {Lozi}, {Guyon}, {Baudoz},
  {Jovanovic}, {Martinache}, {Kudo}, {Serabyn}, \&
  {Kuhn}}]{2015PASP..127..857S}
{Singh}, G., {Lozi}, J., {Guyon}, O., {Baudoz}, P., {Jovanovic}, N.,
  {Martinache}, F., {Kudo}, T., {Serabyn}, E., \& {Kuhn}, J. 2015, \pasp, 127,
  857

\bibitem[{{Singh} {et~al.}(2014){Singh}, {Martinache}, {Baudoz}, {Guyon},
  {Matsuo}, {Jovanovic}, \& {Clergeon}}]{2014PASP..126..586S}
{Singh}, G., {Martinache}, F., {Baudoz}, P., {Guyon}, O., {Matsuo}, T.,
  {Jovanovic}, N., \& {Clergeon}, C. 2014, \pasp, 126, 586

\bibitem[{Tomov {et~al.}(2010{\natexlab{a}})Tomov, Dongarra, \&
  Baboulin}]{tdb10}
Tomov, S., Dongarra, J., \& Baboulin, M. 2010{\natexlab{a}}, Parallel
  Computing, 36, 232

\bibitem[{Tomov {et~al.}(2010{\natexlab{b}})Tomov, Nath, Ltaief, \&
  Dongarra}]{tnld10}
Tomov, S., Nath, R., Ltaief, H., \& Dongarra, J. 2010{\natexlab{b}}, in Proc.
  of the IEEE IPDPS'10 (Atlanta, GA: IEEE Computer Society), 1--8,
  {DOI:~10.1109/IPDPSW.2010.5470941}

\bibitem[{{Trauger} {et~al.}(2016){Trauger}, {Moody}, {Krist}, \&
  {Gordon}}]{2016JATIS...2a1013T}
{Trauger}, J., {Moody}, D., {Krist}, J., \& {Gordon}, B. 2016, Journal of
  Astronomical Telescopes, Instruments, and Systems, 2, 011013

\bibitem[{{Vogt} {et~al.}(2011){Vogt}, {Martinache}, {Guyon}, {Yoshikawa},
  {Yokochi}, {Garrel}, \& {Matsuo}}]{2011PASP..123.1434V}
{Vogt}, F.~P.~A., {Martinache}, F., {Guyon}, O., {Yoshikawa}, T., {Yokochi},
  K., {Garrel}, V., \& {Matsuo}, T. 2011, \pasp, 123, 1434

\end{thebibliography}

\end{document}